\documentclass[conference]{IEEEtran}

\usepackage{amsmath,nicefrac}

\newcommand{\remove}[1]{}

\newcommand{\qed}{\hspace*{\fill}\rule{5pt}{5pt}}

\newtheorem{theorem}{Theorem}[section]
\newtheorem{corollary}[theorem]{Corollary}

\newtheorem{proposition}[theorem]{Proposition}

\newtheorem{definition}[theorem]{Definition}
\newtheorem{remark}[theorem]{Remark}

\newcommand\nc\newcommand
\nc\bfa{{\mathbf a}}\nc\bfA{{\mathbf A}}\nc\cA{{\mathcal A}}
\nc\bfb{{\mathbf b}}\nc\bfB{{\mathbf B}}\nc\cB{{\mathcal B}}
\nc\bfc{{\mathbf c}}\nc\bfC{{\mathbf C}}\nc\cC{{\mathcal C}}
\nc\bfd{{\mathbf d}}\nc\bfD{{\mathbf D}}\nc\cD{{\mathcal D}}
\nc\bfe{{\mathbf e}}\nc\bfE{{\mathbf E}}\nc\cE{{\mathcal E}}
\nc\bff{{\mathbf f}}\nc\bfF{{\mathbf F}}\nc\cF{{\mathcal F}}
\nc\bfg{{\mathbf g}}\nc\bfG{{\mathbf G}}\nc\cG{{\mathcal G}}
\nc\bfh{{\mathbf h}}\nc\bfH{{\mathbf H}}\nc\cH{{\mathcal H}}
\nc\bfi{{\mathbf i}}\nc\bfI{{\mathbf I}}\nc\cI{{\mathcal I}}
\nc\bfj{{\mathbf j}}\nc\bfJ{{\mathbf J}}\nc\cJ{{\mathcal J}}
\nc\bfk{{\mathbf k}}\nc\bfK{{\mathbf K}}\nc\cK{{\mathcal K}}
\nc\bfl{{\mathbf l}}\nc\bfL{{\mathbf L}}\nc\cL{{\mathcal L}}
\nc\bfm{{\mathbf m}}\nc\bfM{{\mathbf M}}\nc\cM{{\mathcal M}}
\nc\bfn{{\mathbf n}}\nc\bfN{{\mathbf N}}\nc\cN{{\mathcal N}}
\nc\bfo{{\mathbf o}}\nc\bfO{{\mathbf O}}\nc\cO{{\mathcal O}}
\nc\bfp{{\mathbf p}}\nc\bfP{{\mathbf P}}\nc\cP{{\mathcal P}}
\nc\bfq{{\mathbf q}}\nc\bfQ{{\mathbf Q}}\nc\cQ{{\mathcal Q}}
\nc\bfr{{\mathbf r}}\nc\bfR{{\mathbf R}}\nc\cR{{\mathcal R}}
\nc\bfs{{\mathbf s}}\nc\bfS{{\mathbf S}}\nc\cS{{\mathcal S}}
\nc\bft{{\mathbf t}}\nc\bfT{{\mathbf T}}\nc\cT{{\mathcal T}}
\nc\bfu{{\mathbf u}}\nc\bfU{{\mathbf U}}\nc\cU{{\mathcal U}}
\nc\bfv{{\mathbf v}}\nc\bfV{{\mathbf V}}\nc\cV{{\mathcal V}}
\nc\bfw{{\mathbf w}}\nc\bfW{{\mathbf W}}\nc\cW{{\mathcal W}}
\nc\bfx{{\mathbf x}}\nc\bfX{{\mathbf X}}\nc\cX{{\mathcal X}}
\nc\bfy{{\mathbf y}}\nc\bfY{{\mathbf Y}}\nc\cY{{\mathcal Y}}
\nc\bfz{{\mathbf z}}\nc\bfZ{{\mathbf Z}}\nc\cZ{{\mathcal Z}}

\nc\bsa{{\boldsymbol a}}\nc\bsA{{\boldsymbol A}}
\nc\bsb{{\boldsymbol b}}\nc\bsB{{\boldsymbol B}}
\nc\bsc{{\boldsymbol c}}\nc\bsC{{\boldsymbol C}}
\nc\bsd{{\boldsymbol d}}\nc\bsD{{\boldsymbol D}}
\nc\bse{{\boldsymbol e}}\nc\bsE{{\boldsymbol E}}
\nc\bsf{{\boldsymbol f}}\nc\bsF{{\boldsymbol F}}
\nc\bsg{{\boldsymbol g}}\nc\bsG{{\boldsymbol G}}
\nc\bsh{{\boldsymbol h}}\nc\bsH{{\boldsymbol H}}
\nc\bsi{{\boldsymbol i}}\nc\bsI{{\boldsymbol I}}
\nc\bsj{{\boldsymbol j}}\nc\bsJ{{\boldsymbol J}}
\nc\bsk{{\boldsymbol k}}\nc\bsK{{\boldsymbol K}}
\nc\bsl{{\boldsymbol l}}\nc\bsL{{\boldsymbol L}}
\nc\bsm{{\boldsymbol m}}\nc\bsM{{\boldsymbol M}}
\nc\bsn{{\boldsymbol n}}\nc\bsN{{\boldsymbol N}}
\nc\bso{{\boldsymbol o}}\nc\bsO{{\boldsymbol O}}
\nc\bsp{{\boldsymbol p}}\nc\bsP{{\boldsymbol P}}
\nc\bsq{{\boldsymbol q}}\nc\bsQ{{\boldsymbol Q}}
\nc\bsr{{\boldsymbol r}}\nc\bsR{{\boldsymbol R}}
\nc\bss{{\boldsymbol s}}\nc\bsS{{\boldsymbol S}}
\nc\bst{{\boldsymbol t}}\nc\bsT{{\boldsymbol T}}
\nc\bsu{{\boldsymbol u}}\nc\bsU{{\boldsymbol U}}
\nc\bsv{{\boldsymbol v}}\nc\bsV{{\boldsymbol V}}
\nc\bsw{{\boldsymbol w}}\nc\bsW{{\boldsymbol W}}
\nc\bsx{{\boldsymbol x}}\nc\bsX{{\boldsymbol X}}
\nc\bsy{{\boldsymbol y}}\nc\bsY{{\boldsymbol Y}}
\nc\bsz{{\boldsymbol z}}\nc\bsZ{{\boldsymbol Z}}
\def\dist{\qopname\relax{no}{dist}}
\def\poly{\qopname\relax{no}{poly}}
\def\supp{\qopname\relax{no}{supp}}
\nc{\triangleq}{\stackrel{\Delta}{=}}
\nc{\veps}{\varepsilon} 

\begin{document}

\title{Randomized Frameproof Codes: Fingerprinting Plus Validation Minus Tracing}
\author{
\authorblockN{N. Prasanth Anthapadmanabhan}
\authorblockA{Dept. of Electrical and Computer Eng. \\
University of Maryland \\
College Park, MD 20742 \\
Email: nagarajp@umd.edu}
\and
\authorblockN{Alexander Barg}
\authorblockA{Dept. of ECE and Inst. for Systems Research \\
University of Maryland \\
College Park, MD 20742 \\
Email: abarg@umd.edu}
}

\maketitle

\begin{abstract}
We propose randomized frameproof codes for content protection,
which arise by studying a variation of the Boneh-Shaw fingerprinting problem.
In the modified system, whenever a user tries to access his fingerprinted
copy, the fingerprint is submitted to a validation 
algorithm to verify that it is indeed permissible before the 
content can be executed. We show an improvement in the achievable rates 
compared to deterministic frameproof codes and traditional fingerprinting codes. 

For coalitions of an arbitrary fixed size, we construct randomized frameproof 
codes which have an $O(n^2)$ complexity validation algorithm and 
probability of error $\exp(-\Omega(n)),$ where $n$ denotes the 
length of the fingerprints. Finally, we present a connection between
linear frameproof codes and minimal vectors for size-2 coalitions.
\end{abstract}

\section{Introduction}\label{sect:fp-intro}
The availability of content (e.g., software, movies, music etc.)
in the digital format, although with many advantages, has the downside
that it is now easy for users to make copies, perform alterations,
and share the content illegally. Thus there is a dire need for protecting
the content against unauthorized redistribution, commonly termed
as {\em piracy.}

In this paper, we consider a variation of the Boneh-Shaw fingerprinting 
scheme \cite{boneh} for content protection. We start with
an informal description of the problem. We will refer to the legal 
content owner as the {\em distributor} and the legitimate license holders as {\em users}. 
The distributor embeds a unique hidden mark, called a {\em fingerprint}, 
which identifies each licensed copy. The fingerprint locations, however, remain
the same for all users. The collection of fingerprints is
called the codebook and the distributor uses some form of randomization
in choosing the codebook. We assume that changes to the 
actual content render it useless, while the fingerprint may be subject
to alterations. This assumption is reasonable, for instance, in 
applications to software fingerprinting.

A single user is unable to pinpoint any of the fingerprint locations.
However, if a set of users, called a {\em coalition of pirates},
compare their copies, they can infer some of the fingerprint locations
by identifying the differences. The coalition now attempts to create a
pirated copy with a forged fingerprint. In order to define 
the coalition's capability in creating the forgery, 
Boneh and Shaw introduced the {\em marking assumption}, which simply
states that the coalition makes changes only in those positions 
where they find a difference (and hence are definitely fingerprint 
locations) as they do not wish to damage the content permanently. 

The objective of the distributor is to trace one of the guilty 
users whenever such a pirated copy is found. The maximum coalition size 
is a parameter of the problem. Such a collection of fingerprints 
together with the tracing algorithm is called a {\em fingerprinting code}.
This problem has been studied in detail in \cite{boneh,bbk,tardos,pbd},
where various constructions and upper bounds have been presented.

Consider now the modified system where each 
time a user accesses his fingerprinted copy,
the fingerprint is validated to verify whether it
is in fact permissible in the codebook being used and the
execution continues only if the validation is successful. This 
limits the forgery possibilities for the pirates at the cost of an additional
validation operation carried out every time a user accesses 
his copy. The idea is that by designing an efficient validation algorithm, 
we do not pay too high a price.

The advantage of this scheme is demonstrated by an improvement 
in the achievable rates compared to traditional fingerprinting codes,
even though the actual property (cf. Definition \ref{def:fp-def})
is not in general weaker than fingerprinting.
In addition, since the pirates are limited to creating only a
valid fingerprint and because we are interested in unique decoding,
there is no additional tracing needed. The
distributor simply accuses the user corresponding to the
fingerprint in the pirated copy as guilty.

In this case, the coalition is successful if it is able to 
forge the fingerprint of an innocent user, thus 
``framing'' him as the pirate. The distributor's objective 
is to design codes for which the probability that this error 
event occurs is small, deriving the name {\em frameproof
codes}. 

In the deterministic case with zero-error probability, 
frameproof codes arise as a special case of {\em separating codes},
which have been studied over many years since being introduced in
\cite{fried69}. For further references on deterministic
frameproof codes and separating codes, we refer the interested
reader to \cite{sag94,coh-sch03, stad-stin-wei,blackburn}.
In order to emphasize the difference that we consider the 
randomized setting, we call our codes {\em randomized
frameproof codes}.

The rest of the paper is organized as follows. In Section 
\ref{sect:fp-def}, we give a formal definition for randomized
frameproof codes. Achievable rates under no restrictions on
validation complexity are presented in Section \ref{sect:fp-lb}.
In Section \ref{sect:fp-lin}, we show the existence of
linear frameproof codes and exhibit a connection to minimal
vectors for size-2 coalitions. Finally, we design a concatenated code with efficient
validation for arbitrary coalition sizes in Section \ref{sect:fp-conc}.

\section{Problem definition}\label{sect:fp-def}

We will use the following notation. Boldface will denote vectors. 
The Hamming distance between vectors $\bsx_1,\bsx_2$ will be
denoted by $\dist(\bsx_1,\bsx_2).$
We also write $s_{\bsz}(\bsx_1,\dots,\bsx_t)$ to denote the number
of $\bsz^T$ columns in the matrix formed with the vectors 
$\bsx_1,\dots,\bsx_t$ as the rows.
For a positive integer $n,$ the shorthand notation
$[n]$ will stand for the set $\{1,\dots,n \}.$ We use 
$h(p):= -p \log_2 p - (1-p) \log_2 (1-p)$ to denote the binary
entropy function and $D(p||q):=p \log_2(p/q) + (1-p) \log_2((1-p)/(1-q))$ 
to denote the information divergence.

Let $\cQ$ be an alphabet (often a field) of finite size $q$ and let 
$M$ be the number of users in the system. Assume that there 
is some ordering of the users and denote their 
set by $[M]$. The fingerprint for each user is of length $n.$ 

Consider the following random experiment.
We have a family of $q$-ary codes $\{C_k, k \in \cK\}$ 
of length $n$ and size $M.$ In particular, here the code $C_k$ refers
to an {\em ordered set} of $M$ codewords. We pick one of the codes
according to the probability distribution 
function $(\pi(k), k \in \cK).$ For brevity, the result of
this random experiment is called a {\em randomized code}
and is denoted by $\cC.$ The {\em rate} of this code
is $R=n^{-1}\log_q M$. We will refer to elements of the 
set $\cK$ as \emph{keys}. Note that the dependence on $n$ has been
suppressed for simplicity. 

The distributor assigns the fingerprints as follows.
He chooses one of the keys, say $k,$ with probability 
$\pi(k),$ and assigns to user $i$ the $i$th codeword of $C_k,$ 
denoted by $C_k(i).$ Following the standard cryptographic 
precept that the adversary knows the system, 
we allow the users to be aware of the family of codes
$\{C_k\}$ and the distribution $\pi(\cdot),$ but the exact key 
choice is kept secret by the distributor. 

The fingerprints are assumed to be distributed within 
the host message in some fixed locations unknown to the users.
Before a user executes his copy, 
his fingerprint is submitted to a \emph{validation} algorithm,
which checks whether the fingerprint is a valid codeword in
the current codebook. The execution continues only if the
validation succeeds.

A \emph{coalition} $U$ of $t$ users is an arbitrary 
$t$-subset of $[M].$ The members of the coalition are commonly
referred to as {\em pirates}. Suppose the collection of fingerprints
assigned to $U,$ namely $C_k(U),$ is $\{\bsx_1,\dots,\bsx_t\}.$
The goal of the pirates is to create a forged fingerprint
different from theirs which is valid under the current key choice.

Coordinate $i$ of the fingerprints is called {\em undetectable} for the
coalition $U$ if $x_{1i}=x_{2i}=\dots=x_{ti}$ and is called 
{\em detectable} otherwise. We assume that the coalition follows
the {\em marking assumption} \cite{boneh} in creating the forgery.

\begin{definition}
The {\em marking assumption} states that for any fingerprint $\bsy$
created by the coalition $U$, $y_i=x_{1i}=x_{2i}=\dots=x_{ti}$ in
every coordinate $i$ that is undetectable.
\end{definition}

In other words, in creating $\bsy$, the pirates can modify only detectable positions.

For a given set of observed fingerprints $\{\bsx_1,\dots,\bsx_t\},$ the set of 
forgeries that can be created by the coalition is called the \emph{envelope}. 
Its definition depends on the exact rule the coalition should follow 
to modify the detectable positions \cite{bbk}:

\begin{itemize}
\item If the coalition is restricted to use only a symbol from their
assigned fingerprints in the detectable positions, we obtain the \emph{%
narrow-sense envelope}: 
\begin{equation}  \label{eqn:intro-nenv}
e(\bsx_1,\dots,\bsx_t)= \{\bsy \in \cQ^n | y_i \in \{x_{1i},\dots,x_{ti}\}, 
\forall i \in [n] \};
\end{equation}
\item If the coalition can use any symbol from the alphabet in the
detectable positions, we obtain the \emph{wide-sense envelope}: 
\begin{equation}  \label{eqn:intro-wenv}
E(\bsx_1,\dots,\bsx_t)= \{\bsy \in \cQ^n | y_i=x_{1i}, 
\forall i \text{ undetectable}\}.
\end{equation}
\end{itemize}

For the binary alphabet, both envelopes are exactly the same. 
In the following, we will use $\cE(\cdot)$ to denote the envelope from any of the 
rules mentioned above.

\begin{definition} \label{def:fp-def}
A randomized code $\cC$ is said to be \emph{$t$-frameproof with $\veps$-error}
if for all $U \subseteq [M]$ such that $|U| \le t,$ it holds that
\begin{equation}
\Pr\{ \cE(\cC(U)) \cap (\cC \backslash \cC(U)) \neq \emptyset \} \le \veps,
\end{equation}
where the probability is taken over the distribution $\pi(\cdot).$
\end{definition}

\begin{remark}
Note that the $t$-frameproof property as defined above is not
in general weaker than the $t$-fingerprinting property, i.e.,
a code which is $t$-fingerprinting with $\veps$-error 
\cite[Definition IV.2]{boneh} is not automatically $t$-frameproof 
with $\veps'$-error, for any $0 \le \veps' <1.$
\end{remark}

A straightforward extension of the fingerprinting definition 
yields a randomized code which satisfies the following 
condition: For any coalition of size at most $t$ and any 
strategy they may use in devising a forgery, the probability 
that the forgery is valid is small. However, this definition 
would trivially enable us to achieve arbitrarily high rates. 
Hence, we use the above (stronger) definition.

\section{Lower bounds for binary frameproof codes} \label{sect:fp-lb}

Let us construct a binary randomized code $\cC$ of length $n$ and size $M=2^{nR}$
as follows. We pick each entry in the $M \times n$ matrix
independently to be 1 with probability $p,$ for some $0 \le p \le 1.$

\begin{theorem} \label{thm:fp-lb}
The randomized code $\cC$ is $t$-frameproof with error probability
decaying exponentially in $n$ for any rate 
\begin{equation} \label{eqn:fp-lb}
R < -p^t \log_2 p - (1-p)^t \log_2 (1-p).
\end{equation}
\end{theorem}

\begin{proof}
For $\gamma >0$, define the set of $t$-tuples of vectors 
$$
\cT_{t,\gamma}:=\left\{ (\bsx_1,\dots,\bsx_t):\begin{array}{l}
s_{\bf 1}(\bsx_1,\dots,\bsx_t) \in I_\gamma,\\
s_{\bf 0}(\bsx_1,\dots,\bsx_t) \in J_\gamma
\end{array} \right\},
$$
where $I_{\gamma}:= [n(p^t-\gamma),n(p^t + \gamma)]$
and $J_{\gamma}:= [n((1-p)^t-\gamma),n((1-p)^t + \gamma)].$
It is clear that for any coalition $U$ of size $t,$ the 
observed fingerprints $(\bsx_1,\dots,\bsx_t)$ belong to $\cT_{t,\gamma}$
with high probability\footnote{We say that an event occurs with 
high probability if the probability that it fails is at most 
$\exp(-cn)$, where $c$ is a positive constant.}. Hence, we will
refer to $\cT_{t,\gamma}$ as the set of {\em typical} 
fingerprints. For any coalition $U$ of size $t$
\begin{align}
& \Pr\{ \cE(\cC(U)) \cap (\cC \backslash \cC(U)) \neq \emptyset \} \notag\\
& \le \Pr\{\cC(U) \notin \cT_{t,\gamma}\} \notag\\
& \qquad + \Pr\{ \exists \bsy \in \cC \backslash \cC(U): 
\bsy \in  \cE(\cC(U)) | \cC(U) \in \cT_{t,\gamma} \}. \label{eqn:fp-lb1}
\end{align}
The first term in the above equation decays exponentially in $n.$
It is left to prove that the second term is also exponentially
decaying for $R$ satisfying (\ref{eqn:fp-lb}).

A codeword in $\cC \backslash \cC(U)$
is a part of $\cE(\cC(U))$ if it contains a 1 (resp. 0) in all 
$s_{\bf 1}(\cC(U))$ (resp. $s_{\bf 0}(\cC(U))$) positions. 
Since $\cC(U) \in \cT_{t,\gamma},$ by taking a union
bound the second term in (\ref{eqn:fp-lb1}) is at most
$$
2^{nR} p^{n(p^t-\gamma)} (1-p)^{n((1-p)^t-\gamma)},
$$
which decays exponentially in $n$ for
$$ R < -(p^t-\gamma) \log_2 p - ((1-p)^t -\gamma) \log_2 (1-p).$$
The proof is completed by taking $\gamma$ to be arbitrarily small.
\end{proof}

The bias $p$ in the construction of $\cC$ can be chosen optimally
for each value of $t.$ Numerical values of the rate thus obtained are
shown in Table \ref{tab:fp-lb}, where they are
compared with the existence bounds for deterministic zero-error
frameproof codes (from \cite{coh-sch03}) and rates of fingerprinting codes
(from \cite{pbd,p-barg}). Observe that there is a factor of $t$ improvement
compared to the rate of deterministic frameproof codes.

\begin{table}[h]
\renewcommand{\arraystretch}{1.3}
\caption{Comparison of rates}
\label{tab:fp-lb}
\centering
\begin{tabular}{|c|c|c|c|}
\hline
 & \multicolumn{3}{|c|}{Rates} \\
\cline{2-4}
$t$ & Randomized & Deterministic &
Fingerprinting \\
 & Frameproof & Frameproof & \\
\hline
2 & 0.5 & 0.2075 & 0.25\\
3 & 0.25 & 0.0693 & 0.0833\\
4 & 0.1392 & 0.04 & 0.0158 \\
5 & 0.1066 & 0.026 & 0.0006 \\
\hline
\end{tabular}
\end{table}

\section{Linear frameproof codes}\label{sect:fp-lin}

Unlike fingerprinting codes, randomized frameproof codes eliminate 
the need for a tracing algorithm, but the fingerprints still 
need to be validated. As the validation algorithm is
executed everytime a user accesses his copy, we require
that this algorithm have an efficient running time.
Although the codes designed in the previous section have
high rates, they come at the price of an $\exp(n)$
complexity validation algorithm. Linear codes are an obvious
first choice in trying to design efficient frameproof codes as
they can be validated in $O(n^2)$ time by simply
verifying the parity-check equations.

\subsection{Linear construction for $t=2$}

We now present a binary linear frameproof code for $t=2$ which
achieves the rate given by Theorem \ref{thm:fp-lb}.
Suppose we have $M=2^{nR}$ users. We construct a randomized
linear code $\cC$ as follows. Pick a random 
$n(1-R)\times n$ parity-check matrix with 
each entry chosen independently to be 0 or 1 with equal 
probability. The corresponding set of binary vectors which 
satisfy the parity-check matrix form a linear code of size
$2^{nR}$ with high probability. Each user is then assigned a unique 
codeword selected uniformly at random from this collection.
In the few cases that the code size exceeds $2^{nR},$ we simply ignore
the remaining codewords during the assignment. However, note that
since the validation algorithm simply verifies the parity-check equations,
it will pronounce the ignored vectors also as valid.

\begin{theorem} \label{thm:fp-lin}
The randomized linear code $\cC$ is $2$-frameproof with 
error probability decaying exponentially in $n$ for any rate
$R<0.5.$
\end{theorem}

\begin{proof}
As in the proof of Theorem \ref{thm:fp-lb}, we begin by defining 
the set of typical pairs of fingerprints. For $\gamma >0,$ define
$$
\cT_{\gamma}:=\Big \{ (\bsx_1,\bsx_2): s_{ij}(\bsx_1,\bsx_2) \in I_\gamma,
\forall i,j \in \{0,1\} \Big \},
$$
where $I_\gamma:=[n(\nicefrac{1}{4}-\gamma),n(\nicefrac{1}{4}+\gamma)].$
For any coalition $U$ of two users
\begin{align}
& \Pr\{ \cE(\cC(U)) \cap (\cC \backslash \cC(U)) \neq \emptyset \} \notag\\
& \le \Pr\{\cC(U) \notin \cT_\gamma \}  + 
\sum_{(\bsx_1,\bsx_2) \in \cT_\gamma} \Pr\{\cC(U)=(\bsx_1,\bsx_2)\} \notag\\
& \quad \times \Pr\{ \exists \bsy \in \cC: \bsy \in \cE(\bsx_1,\bsx_2) \backslash
\{ \bsx_1,\bsx_2 \} | \cC(U)=(\bsx_1,\bsx_2)\}. \notag 
\end{align}
It can be seen that the first term again decays exponentially in $n.$ 
\remove{ This can be shown as follows.
Denote by $C_H$ the linear code corresponding to the parity-check
matrix $H.$
Let $\cS_{C_H}(d)$ denote the number of ordered pairs of codewords at 
distance $d$ apart and let $\cN_{C_H}(d)$ be the number of codewords
of weight $d$ in $C_H.$ We have $\cS_{C_H}(d)=|C_H|\cN_{C_H}(d)$ 
as the code is linear. Then
\begin{align*}
\Pr\{\cC(U) \notin \cT_\gamma \} 
& = \sum_{H:n(1-R) \times n} \Pr(H) \frac{\sum_{d \notin I_\gamma} \cS_{C_H}(d)}
{|C_H| (|C_H|-1)} \\
& = \sum_{d \notin I_\gamma} {\sf E} 
\left[ \frac {\cS_{\cC}(d)} {|\cC|(|\cC|-1)} \right] \\
& = \sum_{d \notin I_\gamma} {\sf E} 
\left[ \frac {\cN_{\cC}(d)} {|\cC|-1} \right] \\
& \le \sum_{d \notin I_\gamma}  
\frac {{\sf E}[\cN_{\cC}(d)]} {2^{nR}-1}, 
\end{align*}
where the last inequality is due to the fact that $|\cC| \ge 2^{nR}.$
Using the well-known fact that 
${\sf E}[\cN_{\cC}(d)]=\binom{n}{d}2^{-n(1-R)} \le 2^{-n(1-R-h(\frac{d}{n}))},$
it is clear that $\Pr\{\cC(U) \notin \cT_\gamma \}$ is
exponentially decaying.} We now consider the term inside the summation 
$$\Pr\{ \exists \bsy \in \cC: \bsy \in \cE(\bsx_1,\bsx_2) \backslash
\{ \bsx_1,\bsx_2 \} | \cC(U)=(\bsx_1,\bsx_2)\}.$$
Observe that for any two binary vectors $(\bsx_1,\bsx_2) \in \cT_\gamma,$ 
the sum $\bsx_1 + \bsx_2 \notin \cE(\bsx_1,\bsx_2)$ and also 
${\bf 0} \notin \cE(\bsx_1,\bsx_2).$ Therefore, every 
vector in $\cE(\bsx_1,\bsx_2)\backslash\{\bsx_1,\bsx_2\}$ is linearly
independent from $\bsx_1,\bsx_2.$ Thus for any 
$\bsy \in \cE(\bsx_1,\bsx_2) \backslash \{ \bsx_1,\bsx_2 \},$
$$\Pr\{\bsy \in \cC | \cC(U)=(\bsx_1,\bsx_2)\}=\Pr\{\bsy \in \cC \}=
2^{-n(1-R)}.$$ 
Since $(\bsx_1,\bsx_2) \in \cT_\gamma,$ 
$|\cE(\bsx_1,\bsx_2)| \le 2^{n(\nicefrac{1}{2}+ 2\gamma)}.$ By taking 
the union bound and $\gamma$ to be arbitrarily small, we obtain the result.
\end{proof}

\subsection{Connection to minimal vectors}

In this subsection, we show a connection between linear 2-frameproof
codes and minimal vectors. We first recall the definition for minimal
vectors (see, for e.g., \cite{ashbarg}). Let $C$ be a $q$-ary
$[n,k]$ linear code. The support of a vector $\bsc \in C$ is given
by $\supp(\bsc)=\{i \in [n]: c_i \neq 0\}.$ We write 
$\bsc' \preceq \bsc$ if $\supp(\bsc') \subseteq \supp(\bsc).$

\begin{definition}
A nonzero vector $\bsc \in C$ is called {\em minimal} if
${\mathbf 0} \neq \bsc' \preceq \bsc$ implies $\bsc'=\alpha \bsc,$
where $\bsc'$ is another code vector and $\alpha$ is a nonzero
constant.
\end{definition}

\begin{proposition} \label{prop:fp-min}
For any $\bsx_1,\bsx_2 \in C,$ $\bsx_1 \neq \bsx_2,$ if
$\bsx_2 - \bsx_1$ is minimal then 
$e(\bsx_1,\bsx_2) \cap (C \backslash \{\bsx_1,\bsx_2\}) = \emptyset.$
If $q=2,$ the converse is also true.
\end{proposition}

\begin{proof}
Consider any $\bsy \in \cQ^n$ and define the translate 
$\bsy':= \bsy - \bsx_1.$ It follows that
\begin{eqnarray}
\bsy \in C & \Leftrightarrow & \bsy' \in C \label{eqn:fp-min1} \\
\bsy \notin \{\bsx_1,\bsx_2\} & \Leftrightarrow & \bsy' \notin 
\{ {\mathbf 0}, \bsx_2-\bsx_1 \}. \label{eqn:fp-min2}
\end{eqnarray}
Furthermore, if $y_i \in \{x_{1i},x_{2i}\}$, then
$y'_i \in \{0,x_{2i}-x_{1i}\}$ for all $i \in [n].$
Therefore,
\begin{align}
\bsy \in e(\bsx_1,\bsx_2) & \Rightarrow \left \{
\begin{array}{l}
\bsy' \preceq \bsx_2-\bsx_1, \\
\bsy' \neq \alpha (\bsx_2-\bsx_1), \forall \alpha \notin \{0,1\}.
\end{array}
\right. \label{eqn:fp-min3}
\end{align}
Using (\ref{eqn:fp-min1}), (\ref{eqn:fp-min2}), (\ref{eqn:fp-min3}),
we obtain that 
$e(\bsx_1,\bsx_2) \cap (C \backslash \{\bsx_1,\bsx_2\}) \neq \emptyset$
implies that $\bsx_2-\bsx_1$ is non-minimal.

For $q=2,$ it is easily seen that the reverse statement also holds
in (\ref{eqn:fp-min3}) and thus the converse is also true.
\end{proof}

Recall the random linear code constructed by generating
a random $n(1-R) \times n$ parity-check matrix in the previous
subsection. With some abuse of notation, let us denote the (unordered) set
of vectors satisfying the random parity-check matrix also by $\cC.$ 
Let $\cM(\cC)$ denote the set of minimal vectors in
$\cC.$ We have the following companion result to Corollary 2.5
in \cite{ashbarg}.

\begin{corollary}
As $n \to \infty,$
\begin{align*}
{\sf E} \left[ \frac{|\cM(\cC)|}{|\cC |} \right]
& = \left \{
\begin{array}{ll}
1, & R < \nicefrac{1}{2} \\
0, & R > \nicefrac{1}{2}
\end{array}
\right.
\end{align*}
\end{corollary}

\begin{proof}
As a consequence of Proposition \ref{prop:fp-min}, for
any two users $\{u_1,u_2\},$ we obtain
\begin{align*}
& \Pr\{ \cE(\cC(u_1,u_2)) \cap (\cC \backslash \cC(u_1,u_2)) \neq \emptyset \} \\
= & \Pr\{ \cC(u_2)-\cC(u_1) \notin \cM(\cC) \} \\
= & 1 - {\sf E} \left [ \frac{|\cM(\cC)|}{|\cC|-1} \right].
\end{align*}
The first part of the result is now true by Theorem \ref{thm:fp-lin}.
We skip the details of the latter part which is easily proved using
Chernoff bounds.
\end{proof}

\subsection{Linear codes for larger $t$}

In the light of Theorem \ref{thm:fp-lin}, a natural
question to ask is whether there exist randomized linear frameproof
codes for $t>2,$ perhaps allowing even a larger alphabet. It turns out that,
just as in the deterministic case, linear frameproof codes do not
always exist in the randomized setting too.

\begin{proposition} \label{prop:fp-linexist}
There do not exist $q$-ary linear $t$-frameproof codes with 
$\veps$-error, $0\le \veps < 1,$ which are secure with
the wide-sense envelope if either $t>q$ or $q>2.$
\end{proposition}

\begin{proof}
Consider a coalition of $q+1$ users. For any
linear code realized from the family where the observed fingerprints
are, say, $\bsx_1,\dots,\bsx_{q+1},$ the forgery 
$\bsy=\bsx_1+\dots+\bsx_{q+1}$ is a part of $E(\bsx_1,\dots,\bsx_{q+1}).$
In addition, it is also a valid fingerprint as the code is linear.
This proves the first part of the proposition.

To prove the second part, consider an alphabet (a field) with $q>2.$
For any two pirates with fingerprints $\bsx_1$ and $\bsx_2,$
the forgery $\bsy=\alpha \bsx_1 + (1-\alpha)\bsx_2,$ where
$\alpha \neq 0,1,$ is a valid codeword (by linearity) and is
also a part of the wide-sense envelope.
\end{proof}

Consequently, in considering linear frameproof codes which are 
wide-sense secure, we are limited to $t=2,q=2.$  

\section{Polynomial-time validation for larger $t$}\label{sect:fp-conc}

Usually, the amount of redundancy needed increases with the
alphabet size in fingerprinting applications. Thus, we are
mainly interested in constructing {\em binary} frameproof codes which
have polynomial-time validation. With the binary alphabet,
there is no distinction between wide-sense and narrow-sense envelopes.
Therefore, there do not exist binary linear frameproof codes for $t>2$
by Proposition \ref{prop:fp-linexist}. In this section, we 
use the idea of code concatenation to 
construct a binary frameproof code with polynomial-time validation.

In the case of deterministic codes, if both the inner and outer codes 
are $t$-frameproof ($(t,1)$-separating) with zero-error, then the concatenated code
is also $t$-frameproof. We will now establish a similar result when
the inner code is a randomized $t$-frameproof code.

Let the outer code $C_{\textrm{out}}$ be a (deterministic) $q$-ary linear 
$[N,K,\Delta]$ code. For each of the $N$ coordinates of the outer code, 
generate an independent instance of a randomized binary code 
$\cC_{\textrm{in}}$ of length $m$ and size $q$ which is 
$t$-frameproof with $\veps$-error. Then the concatenated code 
$\cC$ with outer code $C_{\textrm{out}}$ and inner code 
independent instances of $\cC_{\textrm{in}}$ is a randomized 
binary code of length $n=Nm$ and size $q^K.$

\begin{theorem} \label{thm:fp-conc}
If the relative minimum distance of $C_{\textrm{out}}$ satisfies
\begin{equation} \label{eqn:fp-conc}
\frac{\Delta}{N} \ge 1-\frac{1}{t}(1-\xi)
\end{equation}
and the error probability $\veps < \xi$ for $\cC_{\textrm{in}},$
then the concatenated code $\cC$ is $t$-frameproof with error probability 
$2^{-N D(\xi||\veps)}$ and has a $\poly(n)$ validation algorithm.
\end{theorem}

\begin{proof}
In the proof, all vectors are $q$-ary corresponding to the outer
alphabet. Define 
\begin{align*}
s(\bsy,\{\bsx_1,\dots,\bsx_t\})& := |\{i \in [N]:y_i \in \{x_{1i},\dots,x_{ti}\}\}|, \\
d(\bsy,\{\bsx_1,\dots,\bsx_t\})& := \min_{i \in [t]} \dist(\bsy,\bsx_i).
\end{align*} 
Consider a coalition $U \subseteq \{1,\dots,q^K\}$ of size $t.$
For any coordinate $i \in [N]$ of the outer code, the coalition
observes at most $t$ different symbols of the outer alphabet,
i.e., at most $t$ different codewords of the inner code.
Thus if the $t$-frameproof property holds for the observed 
symbols for the realization of $\cC_{\textrm{in}}$ at coordinate $i,$ then at the
outer level the coalition is restricted to output one of the symbols
it observes, i.e., the narrow-sense rule (\ref{eqn:intro-nenv}) holds. On the other hand,
a failure of the $t$-frameproof property at the inner level 
code implies that the coalition is able to create
a symbol different from what they observe in the corresponding coordinate
at the outer level. 

Accordingly, let $\chi_i, i=1,\dots,N,$ denote the indicator random variables (r.v.s)
for failures at the inner level with $\Pr\{\chi_i=1\} \le \veps$
since the inner code has $\veps$-error. 
Note that $\chi_i$ are independent because we have an independent 
instance of the randomized code for every $i=1,\dots,N.$
Then $Z=\sum_{i=1}^N \chi_i$ is a Binomial r.v. denoting the number
of coordinates where the narrow-sense rule fails at the outer
level. For $0 \le z \le N,$ let $e_z(\cdot)$ denote the envelope when the narrow-sense 
rule is followed only in some $N-z$ outer-level coordinates, i.e.,
$$e_z(\bsx_1,\dots,\bsx_t)=\{\bsy: s(\bsy,\{\bsx_1,\dots,\bsx_t\}) \ge N-z \}.$$
For any $\bsy \in e_z(\bsx_1,\dots,\bsx_t),$ there exists 
some $l \in \{1,\dots,t\}$ such that $s(\bsy,\bsx_l) \ge (N-z)/t,$
i.e., $\dist(\bsy,\bsx_l) \le N - (N-z)/t.$ Therefore, 
\begin{equation} \label{eqn:fp-conc1}
e_z(\bsx_1,\dots,\bsx_t) \subseteq \left\{\bsy: d(\bsy,\{\bsx_1,\dots,\bsx_t\}) 
\le N - \frac{N-z}{t}\right\}.
\end{equation}
The coalition $U$ succeeds when it creates a forgery 
which is valid in the outer code. Thus the probability of error
is at most
\begin{align}
 & \Pr\{\exists \bsy \in C_{\textrm{out}} \backslash C_{\textrm{out}}(U):
\bsy \in e_Z(C_{\textrm{out}}(U)) \} \notag \\
\le & \Pr\left\{ \exists \bsy \in C_{\textrm{out}} \backslash C_{\textrm{out}}(U):
d(\bsy,C_{\textrm{out}}(U)) \le N - \frac{N-Z}{t} \right\} \label{eqn:fp-conc2} \\
= & \Pr \left\{ N - \frac{N-Z}{t} \ge \Delta \right \} \label{eqn:fp-conc3} \\
\le & \Pr\{Z \ge N \xi \} \label{eqn:fp-conc4} \\
\le & 2^{-N D(\xi||\veps)}, \label{eqn:fp-conc5}
\end{align}
where (\ref{eqn:fp-conc2}) follows from (\ref{eqn:fp-conc1}), 
(\ref{eqn:fp-conc3}) is because $C_{\textrm{out}}$ is a linear code
with minimum distance $\Delta,$ (\ref{eqn:fp-conc4})
is due to the condition (\ref{eqn:fp-conc}) and
(\ref{eqn:fp-conc5}) is obtained by standard large deviation bounds.

The validation algorithm operates in two steps. In the first
step, the inner code is decoded/validated for every outer code
coordinate by exhaustive search over $q$ codewords.
We then check whether the resulting $q$-ary vector is a member of
the outer code by verifying the parity-check equations.
The claim about the polynomial-time complexity is true by
choosing an appropriate scaling for the inner code length,
for instance, $m \sim \log_2(n).$
\end{proof}

We now make specific choices for the outer and inner codes
in Theorem \ref{thm:fp-conc} to arrive at explicit constructions.
We take $\cC_{\textrm{in}}$ to be the binary randomized $t$-frameproof
code presented in Theorem \ref{thm:fp-lb} and with growing length.
Thus we have the inner code rate as
$$R_{t}= \max_{p \in [0,1]} \left[ -p^t \log_2 p - 
(1-p)^t \log_2 (1-p) \right] $$
and error probability $\veps = 2^{-m \beta}$ for some $\beta >0.$
The outer code $C_{\textrm{out}}$ is a $[q-1,K]$ Reed-Solomon 
(RS) code with rate at most $(1-\xi)/t$ in order to 
satisfy the condition (\ref{eqn:fp-conc}) on the minimum distance.
Observe that for $\veps$ approaching 0 (for large $m$) and $\xi$ fixed, 
$D(\xi||\veps) \sim \xi \log_2(1/\veps).$
Therefore, with $\veps = 2^{-m \beta}$, the error probability 
of the concatenated code is at most $2^{-n(\xi \beta + o(1))}.$
By taking $\xi$ arbitrarily small and $m$ sufficiently large
to satisfy $\veps < \xi$, we obtain the following result.

\begin{corollary}
The binary randomized code obtained by concatenating $C_{\textrm{out}}$
and $\cC_{\textrm{in}}$ is $t$-frameproof with error probability $\exp(-\Omega(n)),$
validation complexity $O(n^2)$ and rate arbitrarily close to $R_t/t.$
\end{corollary}

\section{Conclusion}\label{sect:fp-concl}

The question of upper bounds on the rate of randomized frameproof codes
is open.

\section*{Acknowledgments}

The research is supported in part by NSF grants
CCF0515124 and CCF0635271, and by NSA grant H98230-06-1-0044.

\end{document}